\documentclass[twocolumn,floatfix,superscriptaddress,preprintnumbers,nofootinbib,amsmath,amssymb,aps,prb]{revtex4-1}
\usepackage{graphicx} 
\usepackage{booktabs,siunitx}
\usepackage{wasysym}
\usepackage{dcolumn} 
\usepackage{float}

\usepackage{multirow}

\usepackage[table]{xcolor}
\usepackage{epsfig}

\usepackage{lmodern} 
\usepackage[T1]{fontenc}
\usepackage[utf8]{inputenc}
 
\usepackage[english]{babel}

\usepackage{xfrac}
\usepackage[version=4]{mhchem}
\usepackage{chemformula}

    \definecolor{Acaa}{RGB}{19,145,22}
    \definecolor{A21am}{RGB}{11,36,251}
    \definecolor{I4mmm}{RGB}{124,6,15}
    
\usepackage[final=true,
					bookmarks=true,
					pdftoolbar=true,
					pdfmenubar=true,
					pdftitle={My title},
					pdffitwindow=true,
					pdfauthor={Author},
					colorlinks=true,
					linkcolor=blue,
					citecolor=blue,
					filecolor=blue,
					urlcolor=blue]{hyperref}

\usepackage{todonotes}

\usepackage{verbatim}

\begin{document}

\title{Probing \ce{Ca3Ti2O7} crystal structure at the atomic level: Insights from \ce{^{111m}Cd}/\ce{^{111}Cd} PAC spectroscopy and \textit{ab-initio} studies}
\author{P. Rocha-Rodrigues}
\email{pedro.miguel.da.rocha.rodrigues@cern.ch}
\affiliation{IFIMUP, Institute of Physics for Advanced Materials, Nanotechnology and Photonics, Departamento de Física e Astronomia da Faculdade de Ciências da Universidade do Porto, Rua do Campo Alegre, 687, 4169-007 Porto, Portugal}
\author{I. P. Miranda}
\affiliation{Department of Physics and Astronomy, Uppsala University, Box 516, SE-75120 Uppsala, Sweden}
\author{S. S. M. Santos}
\affiliation{Instituto de Física, Universidade de São Paulo, CP 66318, 05315-970, São Paulo-SP, Brazil}
\author{G. N. P. Oliveira}
\affiliation{IFIMUP, Institute of Physics for Advanced Materials, Nanotechnology and Photonics, Departamento de Física e Astronomia da Faculdade de Ciências da Universidade do Porto, Rua do Campo Alegre, 687, 4169-007 Porto, Portugal}
\author{T. Leal}
\affiliation{IFIMUP, Institute of Physics for Advanced Materials, Nanotechnology and Photonics, Departamento de Física e Astronomia da Faculdade de Ciências da Universidade do Porto, Rua do Campo Alegre, 687, 4169-007 Porto, Portugal}
\author{M. L. Marcondes}
\affiliation{Instituto de Física, Universidade de São Paulo, CP 66318, 05315-970, São Paulo-SP, Brazil}
\author{J. G. Correia}
\affiliation{C$^2$TN, Centro de Ciências e Tecnologias Nucleares, Departamento de Engenharia e Ciências Nucleares, Instituto Superior Técnico, Universidade de Lisboa, Estrada Nacional 10, 2695-066 Bobadela LRS, Portugal}
\author{L. V. C. Assali}
\affiliation{Instituto de Física, Universidade de São Paulo, CP 66318, 05315-970, São Paulo-SP, Brazil}
\author{H. M. Petrilli}
\affiliation{Instituto de Física, Universidade de São Paulo, CP 66318, 05315-970, São Paulo-SP, Brazil}
\author{J. P. Araújo}
\affiliation{IFIMUP, Institute of Physics for Advanced Materials, Nanotechnology and Photonics, Departamento de Física e Astronomia da Faculdade de Ciências da Universidade do Porto, Rua do Campo Alegre, 687, 4169-007 Porto, Portugal}
\author{A. M. L. Lopes}
\email{armandina.lima.lopes@cern.ch}
\affiliation{IFIMUP, Institute of Physics for Advanced Materials, Nanotechnology and Photonics, Departamento de Física e Astronomia da Faculdade de Ciências da Universidade do Porto, Rua do Campo Alegre, 687, 4169-007 Porto, Portugal}
\date{\today}
%
\begin{abstract}
Perturbed angular correlation spectroscopy combined with \textit{ab-initio} electronic structure calculations is used to unravel the structural phase transition path from the low-temperature polar structure to the high-temperature structural phase in  \ch{Ca3Ti2O7}, a hybrid improper ferroelectric. This procedure explores the unique features of a local probe environment approach by monitoring the evolution of the electric field gradient tensor at the calcium 
sites. The local environments, observed above 1057 K, confirm a structural phase transition from the $A2_1am$ symmetry to an orthorhombic $Acaa$ symmetry in the \ch{Ca3Ti2O7} crystal lattice, disagreeing with the frequently reported avalanche structural transition from the polar $A2_1am$ phase to the aristotype $I4/mmm$ phase. Moreover, the EFG temperature dependency, within the $A2_1am$ temperature stability, is shown to be sensitive to the recently proposed \ch{Ca3Ti2O7} ferroelectric polarization decrease within the 500-800~K temperature range.

\end{abstract}

\maketitle

\section{Introduction}
\label{Sec:Intro}

The Ruddlesden-Popper (RP) \ch{Ca3Ti2O7} compound was predicted by Benedek and Fennie, in 2011,  to be a prototypical hybrid improper ferroelectric system through first-principles density functional theory (DFT) calculations~\cite{Benedek_PRL2011}. The experimental confirmation came later, in 2015,  from Oh {\it et al.} research group~\cite{Oh_NM2015a}, that observed hybrid improper ferroelectricity (HIF) in \ch{Ca3Ti2O7} single crystals, by measuring a $\sim$8~$\mu$C/cm$^2$ ferroelectric polarization. 
The discovery of HIF in this naturally layered perovskite (NLP) has prompted the scientific community to revisit other NLP systems
and investigate the nature of structural transitions that they undergo in response to external stimuli, such as temperature or pressure~\cite{Harris2011,Ye2018,Senn_PRL2015,Kratochvilova2019,Pomiro2020,Silva2021,Barbosa2023}.

\begin{figure}[h]
	\centering
	\includegraphics[width=0.8\columnwidth]{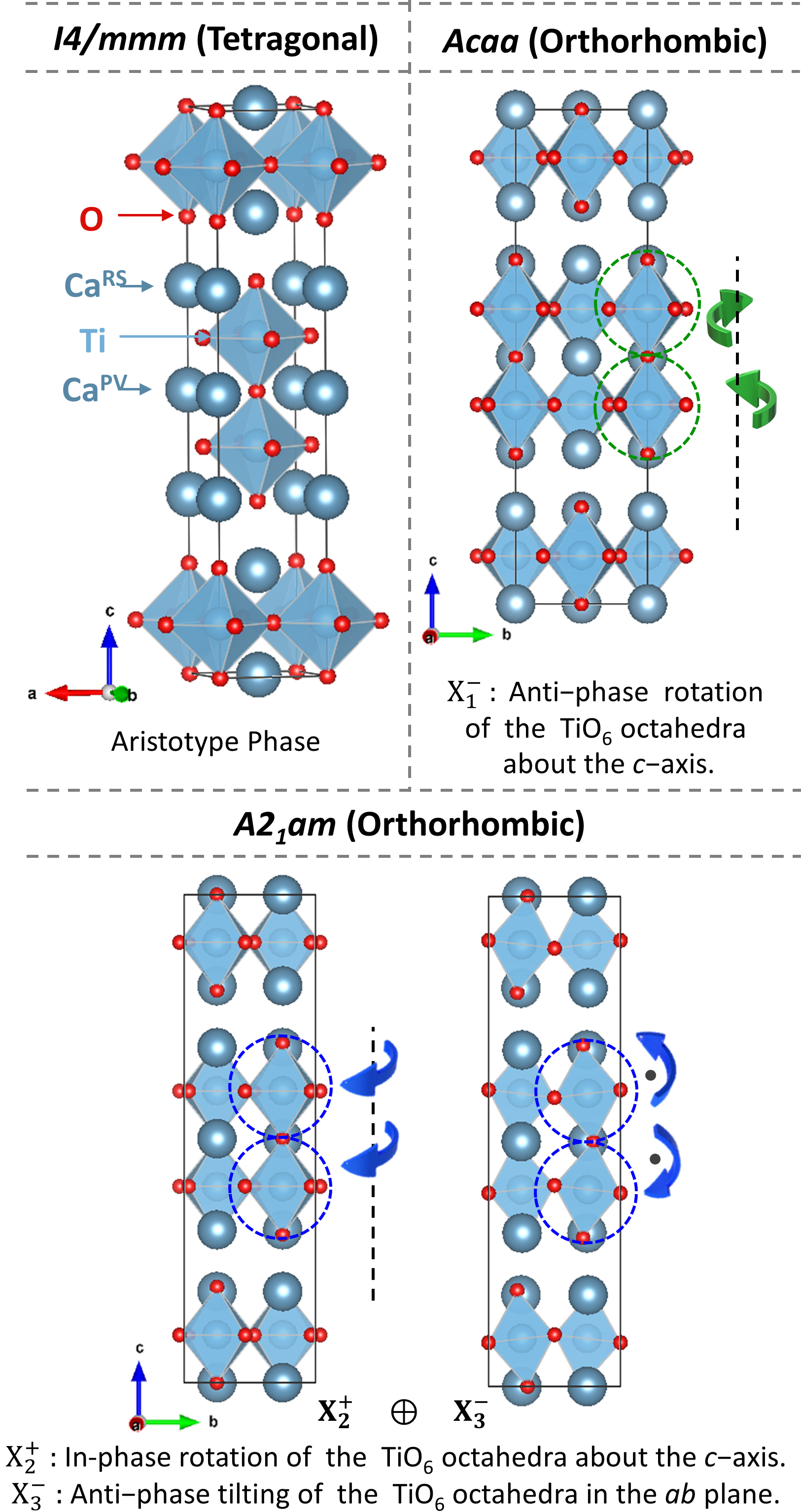}
	\caption{Representation of the \ch{Ca3Ti2O7} crystal structure for the aristotype tetragonal $I4/mmm$ symmetry and of the octahedral rotation and tilting modes for the orthorhombic $Acaa$ and $A2_1am$ space groups. The crystal structure was drawn using VESTA.\cite{Momma2011}}
	\label{fig:CTO_structures}
\end{figure}

The structural phase transition sequence that \ch{Ca3Ti2O7} follows to reach the polar symmetry, as the temperature decreases, remains controversial. The \ch{Ca3Ti2O7} crystal has been assumed to undergo an avalanche structural transition from the high-symmetry $I4/mmm$ phase to the low-symmetry polar $A2_1am$ phase through two distinct rotation and tilting modes of the \ch{TiO6} octahedra that condense simultaneously upon cooling from high temperatures~\cite{Liu2015,Gao_APL2017}. These distortion modes are depicted  in Fig.~\hyperref[fig:CTO_structures]{\ref{fig:CTO_structures}}. Using differential scanning calorimetry (DSC), Liu {\it et al.}~\cite{Liu2015} revealed a first-order structural transition associated with the detection of an endothermic peak at 1099.5~K and an exothermic peak at 1082.5~K, measured for warming and cooling steps, respectively. Moreover, Gao \textit{et al.}~\cite{Gao_APL2017} estimated, from electrical resistivity measurements performed on a \ch{Ca3Ti2O7} single crystal, that the transition between the undistorted $I4/mmm$ and the distorted $A2_1am$ structures should occur around $1063$~K upon warming, and $1039$~K, when cooling the samples.
\begin{figure}[t!]
	\centering
	\includegraphics[width=0.96\columnwidth]{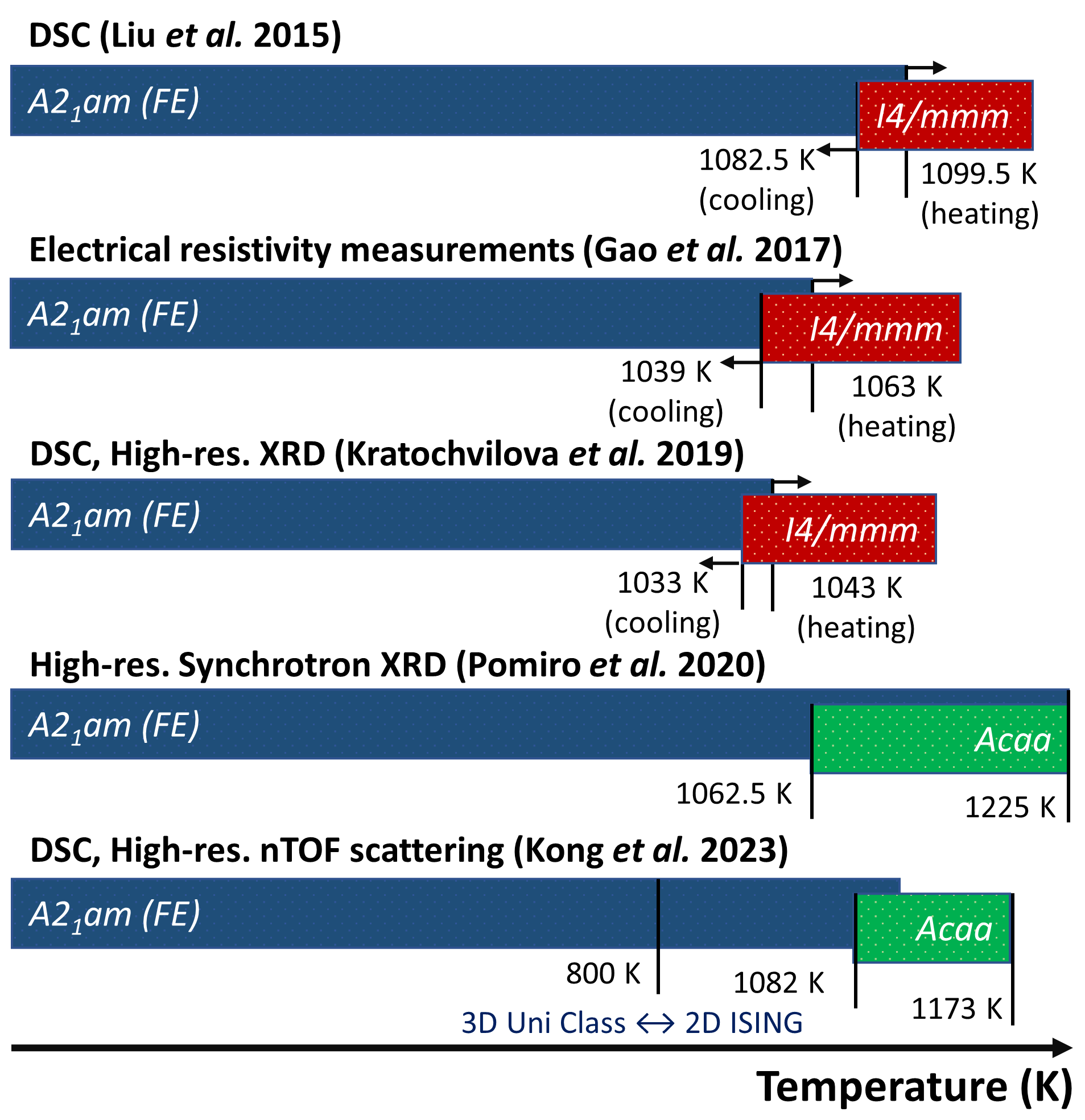}
	\caption{Summary of the \ch{Ca3Ti2O7} high-temperature structural transitions, as reported in a few of the selected works~\cite{Liu2015,Gao_APL2017,Kratochvilova2019,Pomiro2020,Kong2023}.}
	\label{fig:CTO_literature}
\end{figure}
Kratochvilova {\it et al.}~\cite{Kratochvilova2019} conducted synchrotron {X-ray} powder diffraction (XRD) measurements on the Ca$_{3-x}$Sr$_{x}$Ti$_2$O$_7$ series and reported that for $x<0.6$ a direct structural transition from the ferroelectric $A2_1am$ phase to the paraelectric undistorted $I4/mmm$ is observed. 
More recently, Pomiro \textit{et al.}~\cite{Pomiro2020} claimed that the \ch{Ca3Ti2O7} undergoes a first-order structural transition from the $A2_1am$ to the $Acaa$ orthorhombic phase, rather than to the aristotype $I4/mmm$ structure,  
suggesting that the \ch{Ca3Ti2O7} system should follow the same structural transition path as the \ch{Ca3Mn2O7} isostructural compound~\cite{Senn_PRL2015, RRodrigues_PRB2020}. The coexistence of the $A2_1am$ and $Acaa$ structural phases within the $1062.5$~K to $1225$~K temperature range was proposed~\cite{Pomiro2020}. Conversely, Senn \textit{et al.}~\cite{Senn_PRL2015} did not report any structural transition in their XRD measurements on polycrystalline  \ch{Ca3Ti2O7} samples, although they proposed that the samples undergo decomposition around $1150$~K. More recently, using high-resolution neutron time-of-flight (nTOF) scattering combined with DSC experiments, Kong \textit{et al.}~\cite{Kong2023} reported that \ch{Ca3Ti2O7} displays the $Acaa$ symmetry for temperatures above 1082~K. However, unlike Pomiro \textit{et al.} that reported a broad temperature $A2_1am$ and $Acaa$ coexistence, at 1173~K solely the $Acaa$ structural phase was reported to persist. The analysis of temperature-dependent evolution of \ch{Ca3Ti2O7} ferroelectric polarization, as estimated from the Rietveld refinements, allowed the identification of a second critical temperature at 800~K. Below 800~K, Kong \textit{et al.} proposed that the polarization behavior of \ch{Ca3Ti2O7} transitions from a two-dimensional Ising system to a three-dimensional universality class, where the polarization suffers a significant increase while cooling in the 500-800~K range~\cite{Kong2023}. A summary of the high-temperature structural transitions proposed for \ch{Ca3Ti2O7}, along with the respective critical temperatures, is displayed in Fig.~\ref{fig:CTO_literature}. 

The combined use of nanoscopic scale nuclear spectroscopy techniques, such as $\gamma$-$\gamma$ Perturbed Angular Correlation (PAC), and {\it ab-initio} calculations has proven effective for probing local atomic phenomena. This approach elucidates the atomic environment by monitoring the electric field gradient (EFG) tensor at specific nuclear sites, as demonstrated in recent studies~\cite{Santos2021,RRodrigues_PRB2020, Rocha-Rodrigues2020b,Goliveira_SR2020,Barbosa2019}. Specifically, in homologous Ruddlesden-Popper (RP) structures, the analysis of EFG distributions at the Ca sites across distinct temperatures, using radioactive \ch{^{111m}Cd} isotopes as probes, has provided insights into the octahedral rotation modes that underlie the materials' structural phase transitions and their functional properties~\cite{RRodrigues_PRB2020, Rocha-Rodrigues2020b}. 

In the present study, we explore the structural phase transition pathways of the \ce{Ca3Ti2O7} compound. PAC investigations were conducted over a broad temperature range, from $10-1200$~K, employing \ch{^{111m}Cd} nuclei as probes. The experimental observations were combined with simulations within the DFT framework and the supercell scheme for  Cd ion substitution. 

\section{Methodology}
\label{Sec:Methodology}

\subsection{Experimental Details}
\label{SubSec:ExpDetails}
Polycrystalline samples of \ce{Ca3Ti2O7} were synthesized using the solid-state reaction method. Stoichiometric amounts of \ce{CaCO3} (99.0\%, Sigma-Aldrich) and \ce{TiO2} (99.99\%, Sigma-Aldrich) were mixed and subjected to calcination at 1123 K for 3 hours in air. This was followed by a series of grinding, pelletizing, and annealing at 1773 K for 6 hours, repeated several times to ensure homogeneity and phase purity. The phase composition was verified through Le Bail refinement of X-ray powder diffraction data, collected using a Rigaku SmartLab diffractometer and analyzed with the Fullprof software package~\cite{Carvajal_PB1993a}, as detailed in Appendix \ref{SubSec:Crystallographic_data}.

Given the short lifetime of 48.5 min of the \ce{^{111m}Cd} parent probe, performing a PAC measurement at each temperature required an implantation-annealing-measurement cycle at the ISOLDE-CERN facility~\cite{Schell2017}. Experiments were conducted on approximately 4~\ce{mm^3} samples, all obtained from the synthesized pellet. Ion beam-implantation of \ce{^{111m}Cd} isotopes with an energy of 30~keV, up to a low dose of $10^{11}$ ions/cm$^{2}$. The \ce{^{111m}Cd} probe should substitute Ca$^{2+}$ in perovskite systems since it has a similar ionic radius and charge, facilitating accurate PAC measurements\cite{Lopes_RC_PRB2006, RRodrigues_PRB2020, Rocha-Rodrigues2020b}. After each implantation step, a 20-minute thermal annealing was performed in air, typically within the $1100-1123$~K temperature range, to eliminate implantation-induced defects. Each PAC measurement was carried out for a $\sim\,$3~h acquisition time, using a six-detector angular correlation setup~\cite{Correia2005}. For experiments conducted below room temperature, the samples were kept under vacuum using a He closed-cycle refrigerator to reach the desired measurement temperatures through cooling. For experiments above room temperature, a high-temperature furnace exposed the samples to an air atmosphere, achieving the aimed measurement temperatures through heating.

The \ch{^{111m}Cd} probes decay to stable \ch{^{111}Cd} through an intermediate state by emitting two consecutive $\gamma$ photons. The half-life of the parent state is 48.5~min, while the intermediate state has a half-life of 84.5~ns~\cite{Nagl2013}. In the absence of hyperfine interaction, the probability of detecting the second $\gamma$ ray ($\gamma_2$) at an angle $\theta$ relative to the direction of the first emitted $\gamma$ ray ($\gamma_1$), exhibits a characteristic anisotropic angular distribution. This distribution can be described by an expansion in Legendre polynomials ($P_k(\cos \theta)$) and the angular correlation coefficients characteristic of the nuclear decay cascade ($A_{kk}(\gamma_1, \gamma_2)$)~\cite{Nagl2013}. In the presence of hyperfine interactions, between the electric quadrupole moment of the nucleus in the intermediate state and the EFG at the nuclear site, the $\gamma$-$\gamma$ anisotropic angular distribution ($W(\theta,t)$) becomes time-dependent and  for polycrystalline systems can be described as\cite{Butz1989}
\begin{equation}
W(\theta, t) = e^{-t/\tau_{_{\!N}}}\sum_{k} A_{kk}(\gamma_1, \gamma_2)P_k(\cos \theta) G_{kk}(t).
\label{eq1}
\end{equation}
Here, $\tau_{_{\!N}}$ is the intermediate state half-life, and $G_{kk}(t)$ is the time-perturbation factor containing the information about the hyperfine interactions affecting the intermediate state of the nuclear decay. By using a six-detector angular correlation apparatus, the time-dependent perturbation can be measured and used to estimate the strength and symmetry of the EFG~\cite{Correia2005}.

The EFG is the second derivative of the Coulomb potential at a given nuclear site and gives information on the local charge distribution, being a second-order traceless symmetric tensor $V_{ij}$,  that can be diagonalized (principal reference frame), with the convention $|V_{zz}|\geq|V_{yy}|\geq|V_{xx}|$.  Accordingly, it is usually described by two parameters, the principal component $V_{zz}$ and the asymmetry parameter $\eta=(V_{xx}-V_{yy})/V_{zz}$~\cite{Schatz_Book1996}. In polycrystalline systems and static electric quadrupole interactions, the time-perturbation factor can be described as a sum of periodic terms according to the expression:
\begin{equation}\label{eq:Gkk_PAC}
G_{kk}(t)=s_{k0}+\sum_{n=1}^{3}s_{kn}\cos(\omega_{n}t)e^{-\delta\omega_{n}t}~,
\end{equation}
where $t$ is the time spent by the nucleus in the \ch{^{111m}Cd} intermediate probing state, i.e., the time interval between the pair of $\gamma_1$ and $\gamma_2$ photons detection, and the $\omega_{n}$ are related to the transition frequencies between the hyperfine levels when the intermediate nuclear state is split by the hyperfine interaction. The 
$^{111m}$Cd probing level is characterized by an $I=5/2$ nuclear spin momentum which is split by the electric quadrupole interaction into three sublevels. Consequently, in the Fourier transforms (FTs) of the $G_{kk}(t)$ functions,  a triplet of frequencies, $\omega_{1}$, $\omega_{2}$, and $\omega_{3}=\omega_{1}+\omega_{2}$, is observed for each non-vanishing EFG distribution present in the system. The damping term in equation (\ref{eq:Gkk_PAC}) can be related to the presence of randomly distributed vacancies, defects, and lattice strains, therefore, the $\omega_{n}$ transition frequencies are characterized by a $\delta$ relative width about a mean value and each local environment can be described by an EFG Lorentzian-like distribution.   

The $\omega_{n}$ transition frequencies and the $s_{kn}$ correspondent amplitudes are related with the fundamental quadrupolar frequency $\omega_{0}$~\cite{Butz1989}. Explicitly, one can define $\omega_n=g_n(\eta)\omega_0$, where $g_n(\eta)$ and  $s_{kn}=s_{kn}(\eta)$ (Eq. \ref{eq:Gkk_PAC}) are known functions of the asymmetry parameter~\cite{Mendoza1977}.
In the particular case of $\eta= 0$, the fundamental quadrupolar frequency matches the lower observable, i.e., $\omega_{1} = \omega_{0}$, being $\omega_{2} =2\omega_{0}$ and $\omega_{3} = 3\omega_{0}$. 
The fundamental quadrupolar frequency, from which we obtain the strength of the EFG, main component $|V_{zz}|$, is defined as
\begin{equation}\label{eq:w0_PAC}
\omega_{0}=\frac{3eQV_{zz}}{2I(2I-1)\hbar}~,
\end{equation}
where $Q$ is the nuclear electric quadrupole moment of the \ch{^{111m}Cd} probe in the intermediate level. For the calculation of $|V_{zz}|$, we have used the well-known  $Q=0.83(13)$ b value\cite{Schatz_Book1996}.
The $R(t)$ anisotropy function obtained experimentally can be approximated as 
\begin{equation}\label{eq:R(t)_PAC}
R(t)~\approx~A_{22}^{eff}\sum_{i}f_{i}G_{22}^{i}(t), 
\end{equation}
where $A_{22}^{eff}$ corresponds to the corrected solid-angle anisotropy coefficient, and the summation over $i$ accounts for the different observed local environments, each weighted by a population factor $f_i$. The experimental $R(t)$ functions were analyzed with exact numerical methods that construct the expected observable by solving the exact characteristic equation of the hyperfine interaction Hamiltonian using NNFIT software~\cite{Correia2005, Barradas1992}.

\subsection{Computational Details}

First-principles calculations within the DFT\cite{Hohenberg_PR1964,Kohn_PR1965} framework were performed using the Quantum ESPRESSO computational package~\cite{Giannozzi_JPCM2017,Giannozzi_JPCM2009}. 
The exchange-correlation potential was treated based on the generalized gradient approximation (GGA) to describe the electronic structure, as proposed by Perdew, Burke, and Ernzerhof (PBE)~\cite{Perdew_PRL1997}. The calculations were carried out with the projected augmented wave (PAW) method \cite{Blochl_PRB1994}, which has shown to be appropriate for describing electric field gradients \cite{Petrilli_PRB1998}. The wave functions and charge densities were expanded in a plane wave basis set with cutoff energies 110~Ry and 600~Ry, respectively.

High-temperature and structural phase transitions were investigated by following the change of the average atomic positions as a function of temperature, as reported in the literature. We used experimental structural data reported for the $A2_1am$ and $Acaa$ space groups from~ Refs.\citenum{Senn_PRL2015}  and \citenum{Pomiro2020} while the aristotype $I4/mmm$ phase was built from the $Acaa$ reported structure by removing the $X_1^-$ octahedral rotation and the orthorhombic lattice deformation using the ISODISTORT web-based tool\cite{Campbell2006,Stokes}. 
The $A2_1am$ ground state properties were also obtained, self-consistently,  by optimizing the structure and atomic positions until a total force smaller than 0.05 eV/\AA\ was achieved \cite{Marcondes2020}. For all of these \ce{Ca3Ti2O7} structures, we compute the EFG components at the non-equivalent Ca-sites, localized in the perovskite blocks (\ce{Ca^{PV}}) or in the rocksalt-like layers (\ce{Ca^{RS}}). The irreducible first Brillouin zone was sampled with a 7$\times$4$\times$4 Monkhorst-Pack (MP) $k$-mesh~\cite{Monkhorst_PRB1976}. In all cases, convergence tests were performed. 
The inclusion of a Cd ion as a probe in the \ce{Ca3Ti2O7} system, replacing a Ca atom in the appropriate position (\ce{Cd^{RS}} or \ce{Cd^{PV}}), was considered by constructing a 192-atom 2$\times$2$\times$2  supercell, based on the primitive RP crystal structure, and the Brillouin zone was sampled by a 4$\times$2$\times$2 MP $k$-mesh. It is worth mentioning that this supercell size already gave converged results as tested using larger 432-atom 3$\times$3$\times$2 supercells.

To consider relaxations and distortions around the Cd impurity, the force criterium was used by analyzing the displacement of Cd and its neighbors inside a given $r=3.9$\,\AA\ $\,$  radius around the impurity, while keeping fixed the atoms outside this region.

\section{Results}

\subsection{Perturbed Angular Correlation Results}
\label{SubSec:PACCTO}

\begin{figure}[t!]
	\centering
	\includegraphics[width=1\columnwidth]{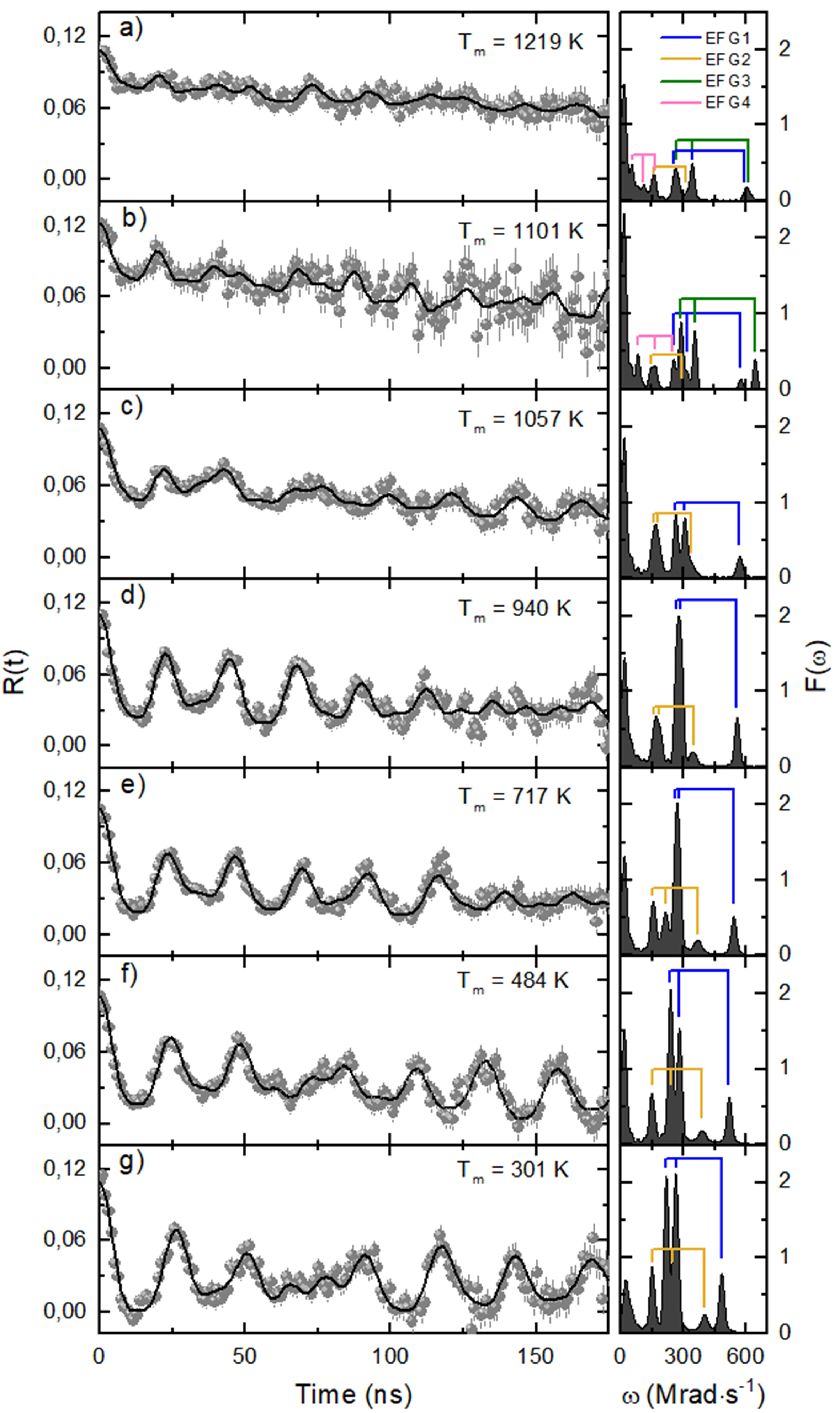}
	\caption{Representative $R(t)$ functions with the corresponding fits, and respective Fourier transforms of the fits taken at different temperatures.}
	\label{fig:CTORT}
\end{figure}

\begin{figure}[t!]
	\centering
	\includegraphics[width=1\columnwidth]{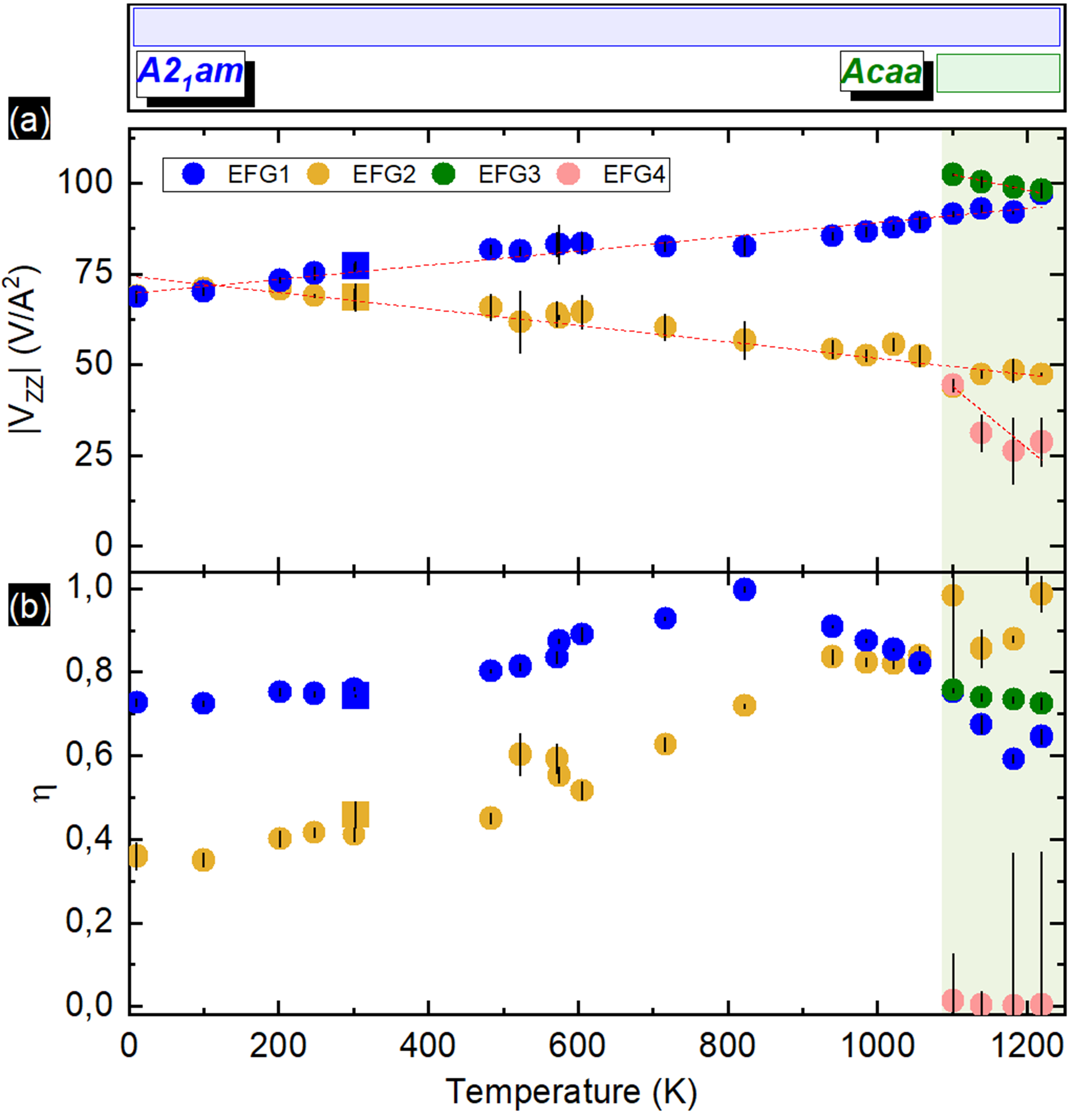}
	\caption{Experimental results  for the \ce{Ca3Ti2O7} sample: \textbf{(a)} EFG strength $|V_{zz}|$ and \textbf{(b)} asymmetry parameter $\eta$. The dashed red line represents the linear least squares fits. The square symbols highlight the fitting parameters obtained when increasing the post-implantation annealing temperature to 1283~K.}
	\label{fig:CTOtrend}
\end{figure}

The experimental perturbation function $R(t)$ and the respective FTs are shown in Fig.~\ref{fig:CTORT}. Throughout the measured temperature range, below 1057~K, it is possible to notice a mixture of at least two EFG distributions, labeled as EFG$_1$ and EFG$_2$. The coexistence of these distributions is highlighted in the FTs graphs by the two frequency triplets identified by the blue and yellow lines. These two fractions represent two well-defined EFG distributions, portraying $R(t)$ profiles with low attenuation (low $\delta$ values). Above the 1057~K temperature, it is possible to refine the experimental $R(t)$ function considering the presence of a third (EFG$_3$) and a fourth (EFG$_4$) distribution, highlighted in the FTs graphs by the green and pink lines.

The thermal dependency of the EFG strength, $|V_{zz}|_i$, and the asymmetry parameter, $\eta_i$, for the different observed local environments ($i = 1$-$4$), are represented in Fig.~\ref{fig:CTOtrend}. At room temperature, the EFG$_1$ distribution, identified as the blue dots in Fig.~\ref{fig:CTOtrend}, is characterized by a $|V_{zz}|_1\approx76$~V/A$^{2}$ with $\eta_1\approx 0.76$, 
while the EFG$_2$ distribution, identified as the yellow dots, is characterized by a $|V_{zz}|_2\approx68$~V/A$^{2}$  with $\eta_2\approx 0.42$.
For temperatures below room temperature, the strength of the EFG$_1$ and EFG$_2$ distributions are identical, but with distinct asymmetry parameters. Upon increasing temperature, above 200~K, the thermal dependency of $|V_{zz}|_1$ and $|V_{zz}|_2$ have opposite trends: while the EFG$_2$ presents the standard behavior, a decrease of the $|V_{zz}|_2$ magnitude for increasing temperature, the EFG$_1$ presents an atypical increase of $|V_{zz}|_1$ magnitude. This latter behavior was also observed when probing the \ce{Ca^{RS}} local environment in the $A2_1am$ structure of the \ch{Ca3Mn2O7} system~\cite{RRodrigues_PRB2020}.

To assess the $|V_{zz}|_i$ trends, one can employ a linear least squares fit, allowing the derivation of a normalized slope defined as
\begin{equation}\label{eq:alpha_trend}
\alpha_{\textit{i}}=\frac{1}{|V_{zz}^0|_i} \frac{\Delta |V_{zz}|_i}{\Delta T}~,
\end{equation}
where $|V_{zz}^0|_i$ corresponds to the intercept obtained from the fit. Table \ref{Tab:Vzz-CTO}, presents the respective $\alpha_{\textit{i}}$ values. An average positive coefficient $\alpha_{\rm 1}= 2.8(2) \times 10^{-4}$~K$^{-1}$ was obtained in the $10$-$1220$~K temperature range for the EFG$_1$, which is of the same order of magnitude of that for the \ce{Ca3Mn2O7} compound in the $A2_1am$ structural phase ($2.3\times 10^{-4}$~K$^{-1}$)\cite{RRodrigues_PRB2020}. 

However, for the \ce{Ca3Ti2O7} system, a close inspection of $|V_{zz}|_1$ trend reveals two peculiar slope changes within the 500-900~K temperature range (highlighted in Fig.~\ref{fig:EFG_vii} of Appendix \ref{SubSec:EFG_Vzztrends}). A slope decrease of $|V_{zz}|_1$ is observed at $\approx$~500~K. This behavior reflects the local structural changes that underlie the gradual \ce{Ca3Ti2O7} ferroelectric polarization (P) decrease, reported to be significant between 500-800~K temperature range\cite{Kong2023}. The second $|V_{zz}|_1$ trend change occurs concomitantly with the increase of the asymmetry parameter towards $\eta_1$ $\approx$~1 ($\approx$~823~K). As highlighted in Appendix \ref{SubSec:EFG_Vzztrends}, where the derived $|V_{yy}|_1$ and $|V_{xx}|_1$ components are also shown, such $|V_{zz}|_1$ trend behavior reflect an EFG principal axes permutation, observed when the magnitude of $|V_{zz}|_1$ and $|V_{yy}|_1$ become identical. This occurs due to the convention $|V_{zz}|\geq|V_{yy}|\geq|V_{xx}|$ without necessarily indicating a change in the crystal's symmetry structure. This phenomenon was already observed in EFG $vs$ P theoretical DFT-based studies, focused on ferroelectric polarization switching paths via changing the amplitude of the \ch{TiO6} rotation or tilting modes of several RP structures~\cite{Santos2021}.

On the other hand, for EFG$_2$, a continuous $|V_{zz}|_2$ decrease is observed, characterized with an average negative coefficient $\alpha_{\rm2}= -3.0(1) \times 10^{-4} $~K$^{-1}$, which is similar to the order of magnitude of the values reported when probing the A-site in perovskite oxide structures~\cite{Lopes_PRL2008}.

In the 11-1219~K temperature range, the \ce{^{111m}Cd} probe distributions average ratio ($f_{\rm{EFG_1}}$:$f_{\rm{EFG_2}}$) is  $\sim$2:1, agreeing with  available $\ce{Ca^{RS}}$ to $\ce{Ca^{PV}}$ ratio of sites distributions in the RP structures. This fact suggests that the Cd probes are well distributed between both positions after the post-implantation annealing procedure.

\begin{table}[b!]
\begin{center}
\caption{Experimental $|V_{zz}|_i$ thermal trends, obtained from linear fits, carried within the 10-1220~K temperature range for the EFG$_1$ and EFG$_2$ distributions, and within 1100-1220~K for the EFG$_3$ and EFG$_4$ distributions.} 
\vspace*{0.2cm}
\begin{tabular}{c|cc} \hline
 & &\\[-3mm]
             &   $\Delta |V_{zz}|_i/\Delta T$              &   $\alpha_{\textit{i}}$\\
EFG$_i$ distribution~~ & & \\[-2mm]
& \\[-3mm]
            &   ~($\times 10^{-2}$ V\AA$^{-2}$K$^{-1}$)~             &   ~($\times 10^{-4}$ K$^{-1}$)~\\ \hline
 & &\\[-3mm]
EFG$_1$                 &  ~1.9$\pm$0.1                          &  ~2.8$\pm$0.2\\
EFG$_2$                 &  -2.2$\pm$0.1
                          & -3.0$\pm$0.1\\
EFG$_3$               &  -4.2$\pm$0.2                          & -2.9$\pm$0.1\\
EFG$_4$               &  -17$\pm$6                         & -7.4$\pm$0.4\\
\hline
\end{tabular}
\label{Tab:Vzz-CTO}
\end{center}
\end{table}

The EFG$_3$ distribution, observed for temperatures higher than $1100$~K, represented as the green spheres in Fig.~\ref{fig:CTOtrend}, is characterized by a $|V_{zz}|_3\approx 102$~V/A$^{2}$ with $\eta_3\approx 0.76$. Furthermore, above 1100~K the strength of the EFG$_3$ distribution decreases with increasing temperature, which is similar to the behavior observed when probing the \ce{Ca^{RS}} local environment in the $Acaa$ \ch{Ca3Mn2O7} structure \cite{RRodrigues_PRB2020}. Within the 1100-1220~K temperature range, an average negative coefficient of $\alpha_{\rm 3}=-2.9(1) \times 10^{-4}$~K$^{-1}$ was obtained. The high value for the asymmetry parameter of the EFG$_3$ distribution is not in agreement with the interpretation of a transition to the aristotype $I4/mmm$ tetragonal phase, but otherwise with a transition to the locally distorted $Acaa$ orthorhombic structure, as the DFT calculations presented in the next section corroborate. The presence of an extra EFG$_4$ distribution could also be observed in this high-temperature range. At 1101 K the  EFG$_4$ distribution is characterized by a magnitude of $|V_{zz}|_4\approx 44$~V/A$^{2}$ with $\eta_4\approx 0$. For higher temperatures, the asymmetry parameter remains low. However, the estimation of the EFG parameters for this distribution was harder to access from the fits, due to the respective lower occupation and the growing contribution of possible local decomposition of the \ce{Ca3Ti2O7} system at high temperatures, as detailed in Appendix \ref{SubSec:extra_phase}. Still, the obtained  ($f_{\rm{EFG_3}}$:$f_{\rm{EFG_4}}$) probe weight distribution average ratio  1.8:1, across the 1101-1219~K temperature range, remains similar to the  2:1 ratio of available $\ce{Ca^{RS}}:\ce{Ca^{PV}}$ sites in the RP structures.

The following DFT calculations focus on the nature of the structural phase transition of the \ce{Ca3Ti2O7} compound, taking into account the effects of Cd ion substitution at various Ca sites in the \ce{Ca3Ti2O7} system under the distinct structural symmetries.

\subsection{DFT calculations}
\label{SubSec:DFT}

\subsubsection{\ch{^{111m}Cd} as a probe within the DFT determined $A2_1am$ ground state structure}
\label{SubSubSec:Probe_location}

Table~\ref{tab:gce-cd3ti2o7-supercell with-relaxation} shows the theoretical EFG results for the substitutional \ce{Cd} at Ca atomic positions: \ce{Cd^{PV}} and \ce{Cd^{RS}},  for the  \ce{Ca3Ti2O7} 192-atom supercell on the polar $A2_{1}am$ ground state phase. 
For comparison, it is also shown the respective EFG values at the \ce{Ca^{PV}} and \ce{Ca^{RS}} sites in the undoped material. One notices that the Ca substitution and the relaxation/distortion introduced by Cd, produce a large variation of the EFG parameters: an increase factor $\sim$2.8 is observed in the PV site, whereas in the RS site, the factor is $\sim$1.8. 
The latter ratio is in good agreement with the expected $\sim$1.9~EFG enhancement due to the polarization of the Cd probe's internal orbitals, arising purely from the ionic lattice in the point charge model~\cite{Schatz_Book1996}. 
 
The results displayed in Table~\ref{tab:gce-cd3ti2o7-supercell with-relaxation} indicate that structural relaxation effects are more important when substitutional \ce{Cd} occupies the Ca \ce{{PV}} site (rather than the \ce{{RS}}), what can be understood by the fact that the \ce{PV} sites have more nearest-neighbors, especially oxygen ligand atoms.

\begin{table}[b!]
\caption{EFG principal component $\lvert V_{zz}\rvert$, in units of V/$\text{\AA}^2$, and asymmetry parameter $\eta$, dimensionless for the theoretically obtained ground state $A2_{1}am$ phase of the \ce{Ca3Ti2O7} 192-atom supercell. The values of $\lvert V_{zz}\rvert$ and  $\eta$ at the \ce{Ca^{PV}} (perovskite) and \ce{Ca^{RS}} (rocksalt) sites are shown, amongst the respective values at the \ce{Cd} site replacing the \ce{Ca} atoms at both nonequivalent positions: \ce{Cd^{PV}} and \ce{Cd^{RS}}.} 
\begin{center}
\label{tab:gce-cd3ti2o7-supercell with-relaxation}
	    \begin{tabular}{l| c c}
 \hline 
 && \\[-2.5mm] 
 Ground state $A2_1am$ structure~~~~ & ~~$\lvert V_{zz}\rvert$~~  & ~~~$\eta $~~~~ \\
  && \\[-2.5mm] \hline
 && \\[-2.5mm] 
\ce{Ca3Ti2O7}:\ce{Cd^{PV}}~    & 61 	& 0.47  \\
 && \\[-2.5mm]
\ce{Ca3Ti2O7}:\ce{Ca^{PV}}~    & 22	& 0.59\\ 
 && \\[-2.5mm]
\ce{Ca3Ti2O7}:\ce{Cd^{RS}}~   &  44	& 0.63 \\
 && \\[-2.5mm]
\ce{Ca3Ti2O7}:\ce{Ca^{RS}}~   & 24	& 0.89\\ \hline
    \end{tabular}
 \end{center}	
\end{table}

We performed an energetic analysis of the substitutional Cd in the $A2_{1}am$ polar phase of the \ce{Ca3Ti2O7} system ground state, using the 192-atom supercell. The total energy difference between the supercells containing either the \ce{Cd^{RS}} or \ce{Cd^{PV}} substitutional impurity, is $\Delta E$$\sim$0.3~eV (per supercell), in favor of the \ce{Ca^{RS}} site.
That $\Delta E$ value is similar to the one found for the \ce{Ca3Mn2O7} compound~\cite{RRodrigues_PRB2020}. However, a fundamental feature differentiates \ce{Ca3Mn2O7} and \ce{Ca3Ti2O7}: while replacement of Ca by Cd in the RS site is favored in the Mn case ~\cite{RRodrigues_PRB2020}, in the Ti-based perovskite, a coexistence of two fractions can be observed in the PAC measurements. These behaviors may be related to distinct energy barriers between the two energy minima of Cd substitutional sites in the Mn- and Ti-based compounds, which might change the migration profiles of the Cd probes post-implantation.

\subsubsection{\ch{^{111m}Cd} as a probe within the experimental determined low temperature $A2_1am$ structure}
\label{SubSubSec:Probe_location_lowT}

Supercell calculations were performed also using the experimental reported $A2_1am$  phase structures \cite{Senn_PRL2015, Pomiro2020},  at the low 125~K temperature. Table~\ref{tab:gce-cd3ti2o7-supercell_high-temperature} shows the DFT theoretical EFG results for the substitutional \ce{Cd^{PV}} and \ce{Cd^{RS}}, and for comparison, the values obtained at the \ce{Ca^{PV}} and \ce{Ca^{RS}} for the pristine (undoped) 125~K structure are also shown. Noticeably, the new value of $\lvert V_{zz}\rvert$ for the \ce{Ca^{RS}}-site is higher (32 V/$\text{\AA}^2$) than the obtained using the determined ground state structure ($24$~V/$\text{\AA}^2$). In comparison, the new value obtained for the \ce{Ca^{PV}}-site (21 V/$\text{\AA}^2$) remains similar to the one obtained in the previously studied structure (22 V/$\text{\AA}^2$). After performing the Ca by Cd substitution for both sites, independently, the strength of the EFG approximately doubles at each case: 21 to 41 V/$\text{\AA}^2$, at the PV site, and  32 to 60 V/$\text{\AA}^2$ at the RS site.  
After performing structural relaxation of Cd probes' nearest-neighbors, for both sites up to a distance of 3.9 $\text{\AA}$, the strength of the \ce{Cd} EFG remains similar at the RS-site, i.e., $\lvert V_{zz}\rvert_{\rm rlx}=64$~V/$\text{\AA}^2$ versus $\lvert V_{zz}\rvert=60$~V/$\text{\AA}^2$, while it increases at \ce{Cd^{PV}} from $\lvert V_{zz}\rvert=41$~V/$\text{\AA}^2$ to $\lvert V_{zz}\rvert_{\rm rlx}=63$~V/$\text{\AA}^2$. These  EFGs obtained at  \ce{Cd^{PV}} and  \ce{Cd^{RS}} sites, after relaxation,  are in very good agreement with the experimental values, as shown in Fig.~\ref{fig:CTOtrend}, where two distinct local environments having identical EFG strengths but distinct asymmetry parameters are observed. 

\begin{table}[b!]
\caption{EFG principal component $\lvert V_{zz}\rvert$, in units of V/$\text{\AA}^2$, and asymmetry parameter $\eta$, dimensionless, for the $A2_{1}am$ phase of the \ce{Ca3Ti2O7} 192-atom supercell, obtained for the structural parameters at 125~K reported  in ref.\citenum{Senn_PRL2015}.
The values at the \ce{Ca^{PV}}, \ce{Ca^{RS}}, \ce{Cd^{PV}}, and \ce{Cd^{RS}} nuclei are shown. The EFG parameters for the Cd nuclei sites are displayed for before ($\lvert V_{zz}\rvert, \eta$) and after ($\lvert V_{zz}\rvert_{\rm rlx}, \eta_{\rm rlx}$)  atomic positions relaxations.}

   \begin{center}
\label{tab:gce-cd3ti2o7-supercell_high-temperature}
	    \begin{tabular}{l| c c | c c}
\hline 
&&& \\[-2.5mm] 
125~K $A2_1am$ structure~~~~~  &  ~$\lvert V_{zz}\rvert$~  & ~~~$\eta$~~~~ &  ~$\lvert V_{zz}\rvert_{\rm rlx}$~  & ~$\eta_{\rm rlx}$~
\\ 
&&& \\[-2.5mm] \hline
&& \\[-2.5mm] 
\ce{Ca3Ti2O7}:\ce{Cd^{PV}}~    & 41 	& 0.99 & 63 	& 0.55   \\
&& \\[-2.5mm] 
\ce{Ca3Ti2O7}:\ce{Ca^{PV}}~    & 21 	& 0.96 & --- 	& --- \\
&& \\[-2.5mm] 
\ce{Ca3Ti2O7}:\ce{Cd^{RS}}~   &  60	& 0.91 & 64 	& 0.76 \\
&& \\[-2.5mm] 
\ce{Ca3Ti2O7}:\ce{Ca^{RS}}~   & 32	& 0.89 & --- 	& --- \\ \hline

    \end{tabular}
 \end{center}	
\end{table}

\subsubsection{EFG parameters for the temperature evolution of the $A2_1am$, $Acaa$, $I4/mmm$ structures for the pristine and Cd-doped compounds}
\label{SubSubSec:Probe_location_Tevol}

\begin{figure}[b!]
\centering
\includegraphics[width=1\linewidth]{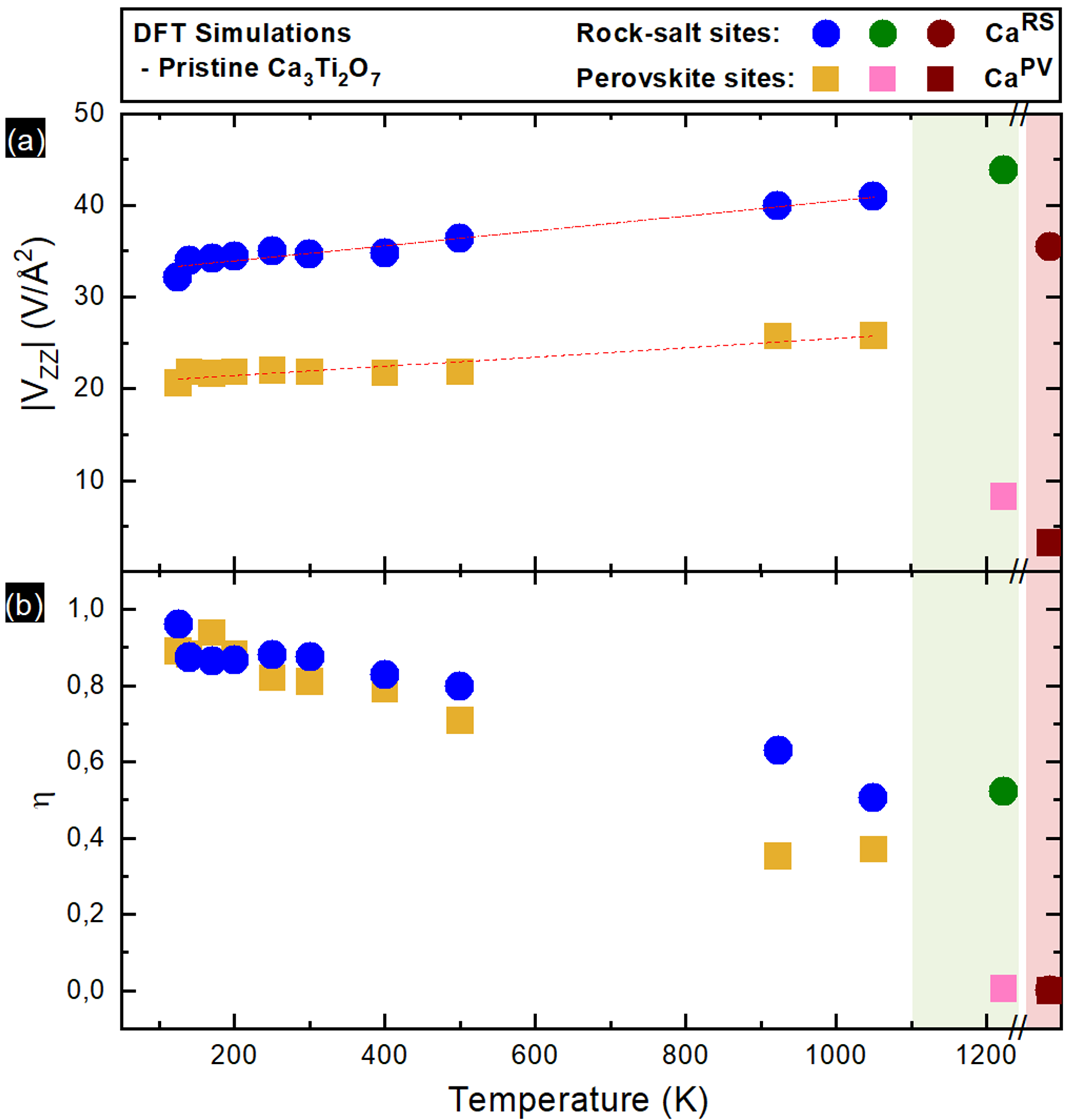}
\caption{EFG parameters, obtained with DFT simulations, in the Ca sites of  pristine \ce{Ca3Ti2O7} systems: \textbf{(a)} Principal component $\lvert V_{zz}\rvert$; \textbf{(b)} asymmetry parameter $\eta$. 
The results are plotted as a function of the corresponding temperature where the diffraction measurements (refs.\citenum{Senn_PRL2015} and \citenum{Pomiro2020}) of the atomic positions, for $A2_1am$ and $Acaa$ phases, used as input in the DFT calculations are performed. For the aristotype $I4/mmm$ phase, the simulated structure was built by removing the lattice and octahedral distortions of the $Acaa$
reported structure through the ISODISTORT tool~\cite{Campbell2006,Stokes}. White, green, and red graphs regions correspond to $A2_1am$, $Acaa$, and  $I4/mmm$ phases, respectively.}
\label{Fig:CTO_PAC_dft_pristine}
\end{figure}

Further DFT studies regarding the EFG parameters for the polar $A2_1am$ phase thermal evolution, as well as for the high-temperature $Acaa$ and $I4/mmm$ structures, were performed. The latter simulations were conducted to clarify the nature of the third and fourth local environments experimentally observed for temperatures higher than 1100~K, as depicted in Fig.~\ref{fig:CTOtrend}.  
Fig.~\ref{Fig:CTO_PAC_dft_pristine} presents the DFT calculated EFG parameters at the \ce{Ca^{PV}} and \ce{Ca^{RS}} sites in the \ce{Ca3Ti2O7} structural phases. 
We calculated the EFGs in the polar $A2_1am$ phase using reported distinct structural lattice and atomic positions obtained for temperatures between 125~K and 1050~K. From linear fitting of the data for the $\lvert V_{zz}\rvert$ trend, the normalized slope values (equation (\ref{eq:alpha_trend})) were extracted and are shown in Table~\ref{Tab:alpha-DFT}.  
A positive $\alpha$ coefficient $\sim$$2.5\times 10^{-4}$~K$^{-1}$ was obtained for both \ce{Ca^{RS}} and \ce{Ca^{PV}} sites, being similar to the experimental value found for the Cd probe  EFG$_1$ distribution (see Table~\ref{Tab:Vzz-CTO}), but disagreeing with the negative trend experimentally observed for the EFG$_2$ distribution. 

\begin{table}[t!]
\begin{center}
\caption{DFT results of the $|V_{zz}|$ thermal trends obtained from linear fitting at the Ca and Cd sites in the polar $A2_1am$ phase, for the pristine and doped supercell structures.}
\begin{tabular}{ccc} \hline
  & & \\[-3mm]
 ~~Atomic~~&   $\Delta |V_{zz}|/\Delta T$              &   $\alpha$\\
   & & \\[-3mm]
   sites   &   ~~($\times 10^{-2}$ V\AA$^{-2}$K$^{-1}$)~~             &   ~~($\times 10^{-4}$ K$^{-1}$) \\ \hline
  & & \\[-3mm]
  \ce{Ca^{RS}}   &  0.81$\pm$0.06                          &  2.5$\pm$0.2\\
    & & \\[-3mm]
  \ce{Ca^{PV}}    &  0.51$\pm$0.06
                          & 2.5$\pm$0.4\\
                            & & \\[-3mm]
\ce{Cd^{RS}}        &  ~4.2& ~7.2\\
  & & \\[-3mm]
\ce{Cd^{PV}}   &  -1.3&  -2.0\\
\hline
\end{tabular}
\label{Tab:alpha-DFT}
\end{center}
\end{table}

\begin{table}[b!]
\caption{EFG principal component $\lvert V_{zz}\rvert$, in units of V/$\text{\AA}^2$, and asymmetry parameter $\eta$, dimensionless, for the $A2_{1}am$ phase of the \ce{Ca3Ti2O7} 192-atom supercell, obtained for the structural parameters at 1050~K reported  in ref.\citenum{Pomiro2020}. The values at the \ce{Ca^{PV}}, \ce{Ca^{RS}}, \ce{Cd^{PV}}, and \ce{Cd^{RS}} nuclei  are shown. The parameters for the Cd nuclei sites are shown before
and after atomic positions relaxation (rlx).}
\begin{center}
\label{tab:gce-cd3ti2o7-supercell_high_923-temperature}
	    \begin{tabular}{l| c c | c c}
\hline 
&&& \\[-2.5mm] 
1050~K $A2_1am$ structure~~  &  ~~$\lvert V_{zz}\rvert$~~  & ~~$\eta$~~~~&  ~~$\lvert V_{zz}\rvert_{\rm rlx}$~  & ~$\eta_{\rm rlx}$~
\\ 
&&& \\[-3mm] \hline
&& &\\[-3mm] 	    
\ce{Ca3Ti2O7}:\ce{Cd^{PV}}~~    & 57 	& 0.27 & 51 	& 0.65   \\
&&& \\[-3mm] 
\ce{Ca3Ti2O7}:\ce{Ca^{PV}}~    & 26	& 0.37  & --- 	& --- \\ 
&&& \\[-3mm] 
\ce{Ca3Ti2O7}:\ce{Cd^{RS}}~   & 81 	& 0.66 &  103 	& 
 0.62   \\
&&& \\[-3mm] 
\ce{Ca3Ti2O7}:\ce{Ca^{RS}}~   &  41	& 0.60  & --- 	& --- \\\hline
    \end{tabular}
 \end{center}	
\end{table}
\begin{figure}[h!]
\centering
\includegraphics[width=1\linewidth]{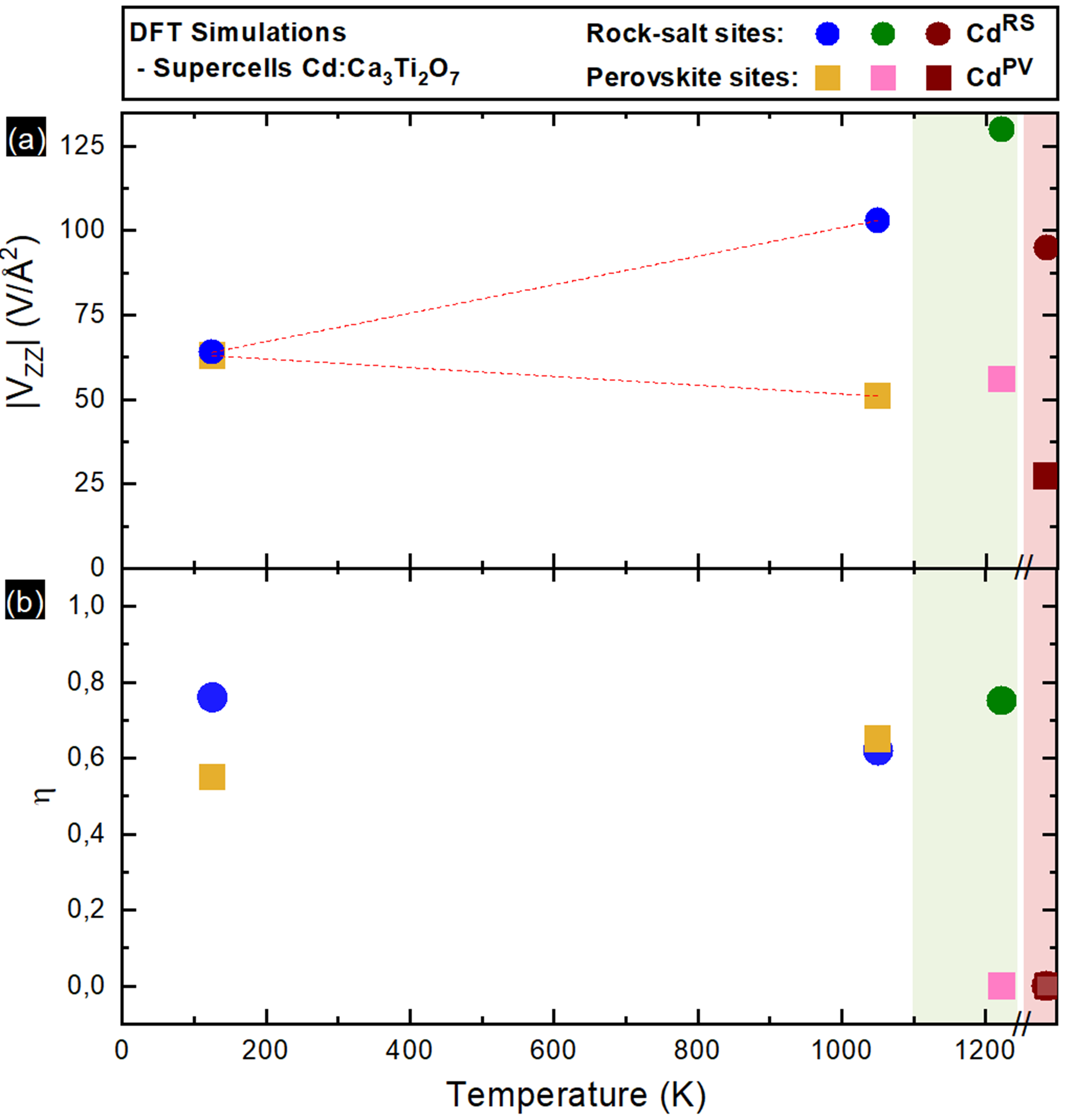}
\caption{EFG parameters, obtained with DFT simulations, in the Cd sites of the Cd-doped \ce{Ca3Ti2O7} compounds: \textbf{(a)} Principal component $\lvert V_{zz}\rvert$; \textbf{(b)} asymmetry parameter $\eta$. The results are plotted as a function of the corresponding temperature where the diffraction measurements (refs.\citenum{Senn_PRL2015} and \citenum{Pomiro2020}) of the atomic positions, for $A2_1am$ and $Acaa$ phases, used as input in the DFT calculations are performed. For the aristotype $I4/mmm$ phase, the simulated structure was built by removing the lattice and octahedral distortions of the $Acaa$ reported structure through the ISODISTORT tool~\cite{Campbell2006,Stokes}. White, green, and red graphs regions correspond to $A2_1am$, $Acaa$, and  $I4/mmm$ phases, respectively.}
\label{Fig:CTO_PAC_dft}
\end{figure}

To address the latter contribution, we have performed calculations, in the supercell scheme, by using as input the 1050~K high-temperature $A2_1am$ phase atomic positions reported in reference \citenum{Pomiro2020}, to investigate the structural relaxation effects when either probing the RS or PV sites with a Cd isotope ion substituting a Ca atom site. After performing structural relaxations, the $\lvert V_{zz}\rvert$ magnitude at the  \ce{Cd^{PV}} site decreases slightly from 57 to 51 V/$\text{\AA}^2$, while at the \ce{Cd^{RS}} site, the EFG strength increases from 81 to 103 V/$\text{\AA}^2$, as can be observed in Table~\ref{tab:gce-cd3ti2o7-supercell_high_923-temperature}.  
As visually depicted in Fig.~\ref{Fig:CTO_PAC_dft} (a), the structural relaxation of Cd, together with its neighboring ions, induces a change in the slope of the trend, as a function of temperature, of $\lvert V_{zz} \rvert$ values at the $A2_1am$ PV site. The results of the supercells calculations show that the effects of structural relaxation around the probe are important for an adequate description of the systems, and that when they are taken into account the $\lvert V_{zz} \rvert$ values in the \ce{Cd^{PV}} and \ce {Cd^{RS}} sites exhibit opposite trends, as observed experimentally.
Furthermore, interestingly, when comparing the results presented in  
figures \ref{fig:CTOtrend} (b) and \ref{Fig:CTO_PAC_dft} (b), we can realize that the interchange of the $\eta$'s values between the EFG$_1$ and EFG$_2$  fractions is captured by the \ce{Cd^{PV} } and \ce{Cd^{RS}} centers when relaxation is considered.

Regarding the pristine \ce{Ca3Ti2O7} high-temperature $Acaa$ and $I4/mmm$ structural phases, the experimental $\eta$ asymmetry parameter is expected to be zero for the PV sites of both phases and for the RS site of the undistorted high symmetry $I4/mmm$ phase, as shown in Fig.~\ref{Fig:CTO_PAC_dft_pristine}. In turn, a higher asymmetry value is expected only for the RS position of the $Acaa$ phase. To confirm such expectations, EFG parameters were also theoretically obtained for the Cd nuclei in the Cd-doped \ce{Ca3Ti2O7} supercell systems, considering the $Acaa$ space group by using the experimental structural data, at 1223~K, reported in Ref. \citenum{Pomiro2020} and the extrapolated $I4/mmm$ structure. The results are shown in Table~\ref{Tab:Vzz-DFT-Acaa}. Indeed, the results show that $\eta$ is only non-zero when the Cd ion occupies the \ce{Ca^{RS}} site of the $Acaa$ structural phase, being $\eta$$\sim$0.6 in the non-optimized case and $\eta$$\sim$0.75 when atomic relaxations were taken into account. These values, when compared to the ones obtained from the PAC experiments, shown in Fig.~\ref{fig:CTOtrend}, corroborate that the EFG$_3$ distribution observed at high temperatures corresponds to the probing of Cd at the \ce{Ca^{RS}} site of the $Acaa$ phase. This observation supports the conclusion that the high-temperature phase transition of \ce{Ca3Ti2O7} involves a first-order structural transition from the polar $A2_1am$ phase to the $Acaa$, in agreement with Pomiro \textit{et al.}~\cite{Pomiro2020}, and similar to what was reported for the \ce{Ca3Mn2O7} system~\cite{Senn_PRL2015,RRodrigues_PRB2020}. This confirms that both $A2_1am$ and $Acaa$ structural phases coexist up to 1219~K and contradicts the commonly reported avalanche structural transition (see figure \ref{fig:CTO_literature}) from the polar $A2_1am$ to the aristotype $I4/mmm$ phase~\cite{Gao_APL2017,Kratochvilova2019,Liu2015}.

\begin{table}[t!] 
\caption{EFG principal component $\lvert V_{zz}\rvert$, in units of V/$\text{\AA}^2$, and asymmetry parameter $\eta$, dimensionless, for the $Acaa$ phase of the \ce{Ca3Ti2O7} 192-atom supercell, obtained for the structural parameters at 1223~K, reported  in ref.\citenum{Pomiro2020}, and for a hypothetical $I4/mmm$ phase. The values at the \ce{Ca^{PV}}  and \ce{Ca^{RS}} sites are shown, amongst the respective values at the \ce{Cd} site replacing the \ce{Ca} atoms at both nonequivalent positions \ce{Cd^{PV}} and \ce{Cd^{RS}}. The EFG parameters for the Cd nuclei sites are shown before and after atomic positions relaxation (rlx).}

   \begin{center}

\color{black}
\label{Tab:Vzz-DFT-Acaa}
	    \begin{tabular}{l| c c | c c}
\hline 
&&& \\[-3mm] 
1223~K $Acaa$ structure~~~  &  ~~$\lvert V_{zz}\rvert$~~  & ~~$\eta$~~~~&  ~~$\lvert V_{zz}\rvert_{\rm rlx}$~~  & ~~$\eta_{\rm rlx}$~~
\\ 
&&& \\[-3mm] \hline
&& &\\[-3mm] 	    
\ce{Ca3Ti2O7}:\ce{Cd^{PV}}~~    & 17 	& 0 & 56 	& 0   \\
&&& \\[-3mm] 
\ce{Ca3Ti2O7}:\ce{Ca^{PV}}~    & 8 & 0 & --- 	& --- \\ 
&&& \\[-3mm] 
\ce{Ca3Ti2O7}:\ce{Cd^{RS}}~   &  87	& 0.60 & 130 	& 0.75   \\
&&& \\[-3mm] 
\ce{Ca3Ti2O7}:\ce{Ca^{RS}}~   &  43	& 0.55 & --- 	& --- \\\hline

&&& \\[-3mm] 
$I4/mmm$ structure  &  ~$\lvert V_{zz}\rvert$~  & ~~$\eta$~~&  ~$\lvert V_{zz}\rvert_{\rm rlx}$~  & ~$\eta_{\rm rlx}$~
\\ 
&&& \\[-3mm] \hline
&& &\\[-3mm] 	    
\ce{Ca3Ti2O7}:\ce{Cd^{PV}}~    & 7 	& 0 & 27 	& 0   \\
&&& \\[-3mm] 
\ce{Ca3Ti2O7}:\ce{Ca^{PV}}~    & 3 & 0 & --- 	& --- \\ 
&&& \\[-3mm] 
\ce{Ca3Ti2O7}:\ce{Cd^{RS}}~   &  74	& 0 & 95 	& 0   \\
&&& \\[-3mm] 
\ce{Ca3Ti2O7}:\ce{Ca^{RS}}~   &  35	& 0 & --- 	& --- \\\hline
    \end{tabular}
 \end{center}	
\end{table}

\section{Conclusions}
\label{Sec:Conclusions}

PAC spectroscopy, combined with {\it ab-initio} calculations, has proven to be a valuable tool for probing the octahedral rotations of the hybrid improper ferroelectric \ce{Ca3Ti2O7}. Local probe experimental data was acquired in a broad temperature range, and {\it ab-initio} calculations were conducted to account for Cd ions substitutionally occupying the Ca sites in the \ce{Ca3Ti2O7} crystal lattice across distinct structural symmetries. The results revealed that Cd probes occupy both Ca rock-salt and Ca perovskite sites in the low temperature $A2_1am$ phase. The EFG temperature dependency at the rock-salt calcium sites, within the $A2_1am$ temperature stability, is also shown to be sensitive to the recently proposed \ch{Ca3Ti2O7} ferroelectric polarization sharp decrease within the 500-800~K temperature range. Furthermore, the analyses indicate that the structural phase transition of the low-temperature polar phase $A2_1am$, as the temperature increases, occurs at a temperature above 1057~K, leading to a crystalline structure of orthorhombic symmetry, which belongs to the space group $Acaa$, contrary to what has often been reported as an avalanche structural transition to the $I4/mmm$ aristotype high-symmetry tetragonal phase. The $A2_1am$ and the $Acaa$ structural phases are shown to coexist up to 1219~K.

\begin{acknowledgments}
\label{Sec:Acknowledgements}

We acknowledge the support of projects CERN/FIS-TEC/0003/2021, NORTE-01-0145-FEDER-022096, POCI-01-0145-FEDER-029454, POCI-01-0145-FEDER-032527, NORTE-01-0145-FEDER-000076, and the Portuguese Foundation for Science and Technology (FCT) under AMLL (2021.04084.CEECIND) and PRR (SFRH/BD/117448/2016) grants. We also acknowledge funding from the European Union’s Horizon 2020 Framework Programme for Research and Innovation under grant agreement no. 654002 (ENSAR2), which supported the IS647 ISOLDE-CERN experiment, and the ISOLDE-CERN collaboration which also financially supported the IS679 experiment. Further, we are grateful for the financial support from the Federal Ministry of Education and Research (BMBF) through grants 05K16PGA and 05K22PGA, as well as support from Brazilian Federal Government Agencies CNPq (grants 314884/2021-1, 308438/2022-1, and 151664/2022-6) and FAPESP (grants 2018/07760-4 and 2022/10095-8).
Additionally, we thank the National Laboratory for Scientific Computing (LNCC/MCTI, Brazil) for providing HPC resources of the Santos Dumont supercomputer (http://sdumont.lncc.br), the Centro Nacional de Processamento de Alto Desempenho em São Paulo (CENAPAD-SP) for their HPC resources, PRACE for awarding us access to Galileu 100 hosted by Cineca, Italy, and the National Academic Infrastructure for Supercomputing in Sweden (NAISS, projects 2023/1-10, 2023/5-454, and 2023/22-1107) at both the National Supercomputing Centre (NSC, Tetralith) and the PDC Centre for High Performance Computing (PDC-KTH, Dardel). IPM would like to express gratitude to H. C. Herper for helpful discussions. Finally, we extend our sincere thanks to all the technical teams at ISOLDE for their exceptional work in delivering high-quality beams for the presented PAC measurements. The collective support from all these entities has been instrumental in achieving the research results reported in this paper.

\end{acknowledgments}


\appendix

\appendix
\label{Sec:Appendix}

\renewcommand\thefigure{\thesection\arabic{figure}} 
\renewcommand\thetable{\thesection\arabic{table}}   
\renewcommand\thesection{\Alph{section}}
\renewcommand\thesubsection{\thesection.\arabic{subsection}}

\section{Crystallographic data of  \ce{Ca3Ti2O7} studied sample at room temperature}
\label{SubSec:Crystallographic_data}
\setcounter{table}{0}
\setcounter{figure}{0}   
\begin{figure}[h!]
	\centering
	\includegraphics[width=1\columnwidth]{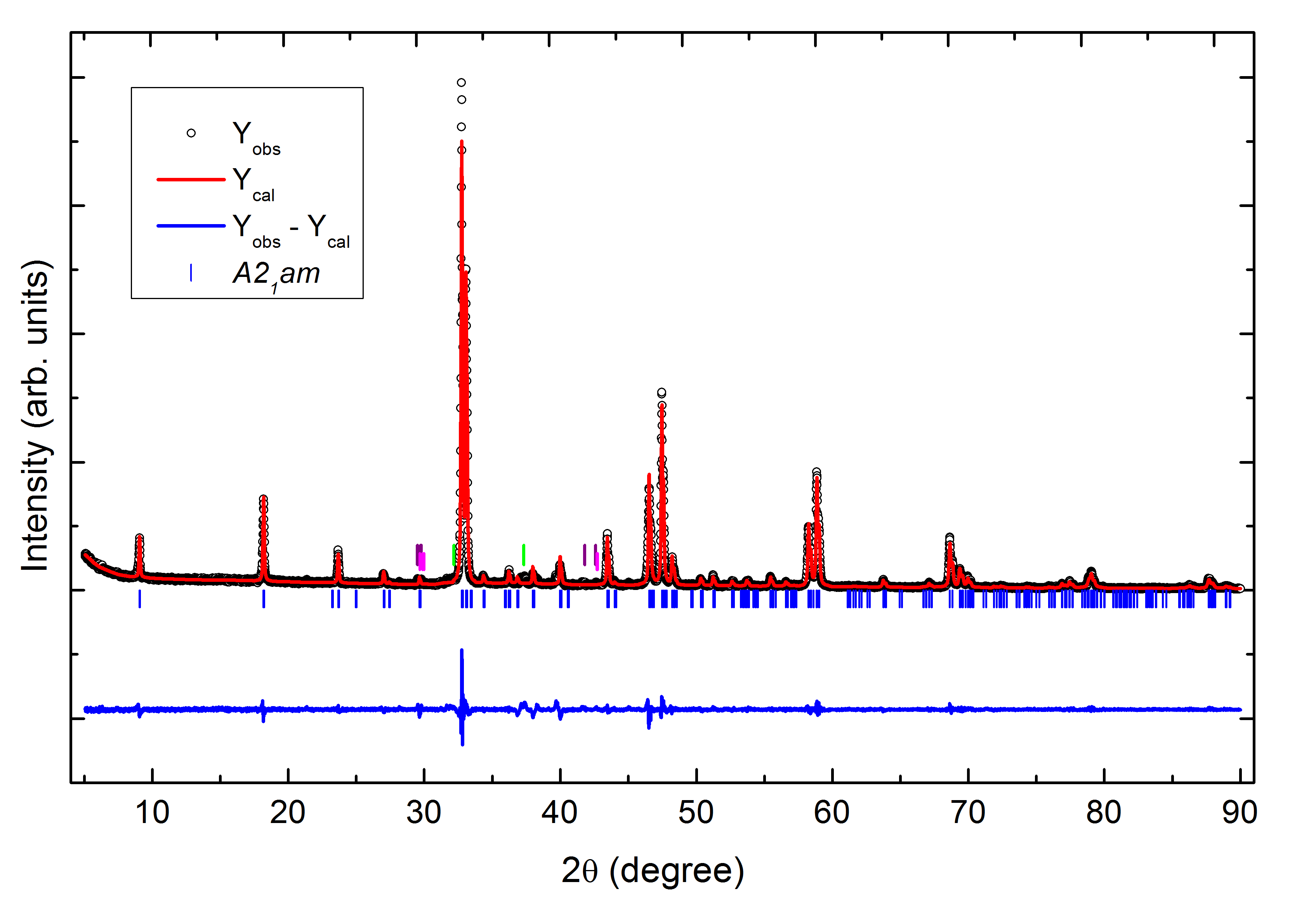}
	\caption{Profile matching fit to data collected by Cu-K$_\alpha$ radiation x-ray diffraction for \ce{Ca3Ti2O7} at room temperature. The intensities of the observed pattern are plotted as black open circles. The calculated pattern is displayed as the red line and the difference between the calculated and measured diffractogram is the blue line. Blue ticks represent Bragg reflections for the \ce{Ca3Ti2O7} $A2_1am$ spacegroup, and the purple ticks represent Bragg reflections for nonabsorbed Cu-K$_\beta$ radiation for the $A2_1am$ spacegroup as well. Pink and green ticks highlight the Bragg peaks related to a minor presence of intermediary \ce{CaTiO3} and \ce{CaO} compounds.}
	\label{fig:CTOXray}
\end{figure}

The synthesized sample's crystallographic structure and lattice parameters have been checked and refined employing XRD at room temperature,  as shown in Fig.~\ref{fig:CTOXray},  where profile matching analysis was performed using the refinement Fullprof software package~\cite{Carvajal_PB1993a}. 
An $A2_{1}am$ structural phase was considered to fit the diffraction pattern, achieving a $\chi^2$ = 2.12 fit-goodness. The crystallographic parameters obtained from the refinement of the \ch{Ca3Ti2O7} $A2_1am$ structure are: $a = 5.4156(2)$ \AA, $b = 5.4126(2)$ \AA, and $c = 19.5067(4)$ \AA.

\section{EFG$_1$ distribution: $\lvert V_{zz} \rvert_1$, $\lvert V_{yy} \rvert_1$ , $\lvert V_{xx} \rvert_1$ thermal trends (Probing $A2_1am$ phase at the Calcium rock-salt site)}
\label{SubSec:EFG_Vzztrends}
\setcounter{table}{0}
\setcounter{figure}{0}

\begin{figure}[h!]
	\centering
	\includegraphics[width=1\columnwidth]{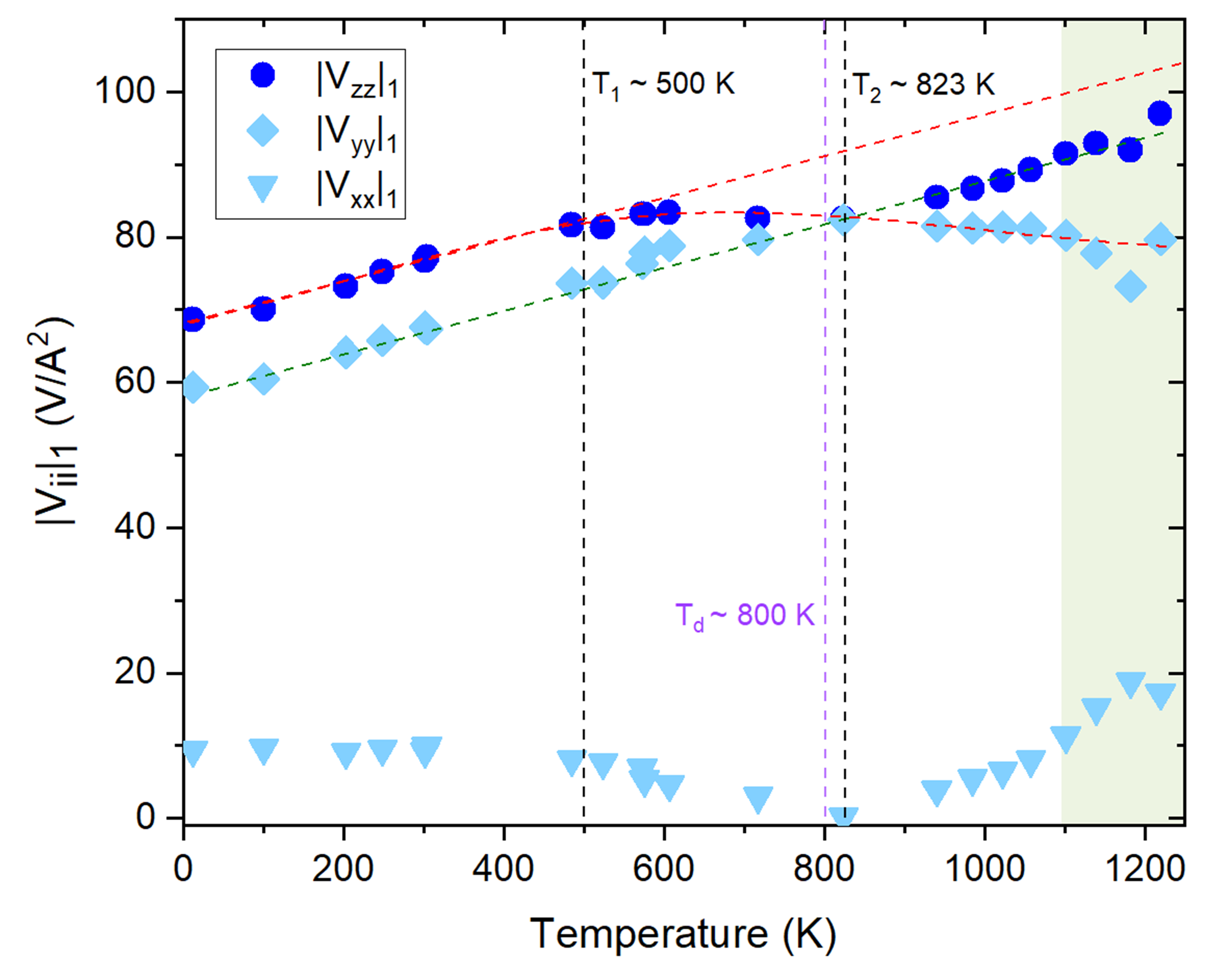}
	\caption{Experimental thermal trends within the 10-1220~K temperature range for $|V_{zz}|_1$ and derived $|V_{yy}|_1$, $|V_{xx}|_1$. T$_1$ and T$_2$ delimit the temperature window where slope changes in $|V_{zz}|_1$(T) are observed. Dashed red and green lines atop the experimental data serve as visual guides to highlight these trend changes in $|V_{zz}|_1$ and $|V_{yy}|_1$. T$_1$~$\approx$~500~K highlights gradual local structure changes following the reported decrease in \ch{Ca3Ti2O7} ferroelectric polarization, where it was reported that at T$_d$~$\approx$~800~K the polarization behavior of \ch{Ca3Ti2O7} eventually transitions from a three-dimensional universality class to a two-dimensional Ising model~\cite{Kong2023}. While T$_2$~$\approx$~823~K marks an EFG principal axes permutation when the magnitude of $|V_{zz}|_1$ and $|V_{yy}|_1$ become identical. This occurs due to the convention $|V_{zz}|\geq|V_{yy}|\geq|V_{xx}|$ without necessarily indicating a change in the crystal's symmetry structure.}
	\label{fig:EFG_vii}
\end{figure}

\section{Low strength and broad EFG distribution induced by high-temperature annealing/measurement conditions \\ in the \ce{Ca3Ti2O7} compound}
\label{SubSec:extra_phase}
\setcounter{table}{0}
\setcounter{figure}{0}

\begin{figure}[b!]
	\centering
	\includegraphics[width=1\columnwidth]{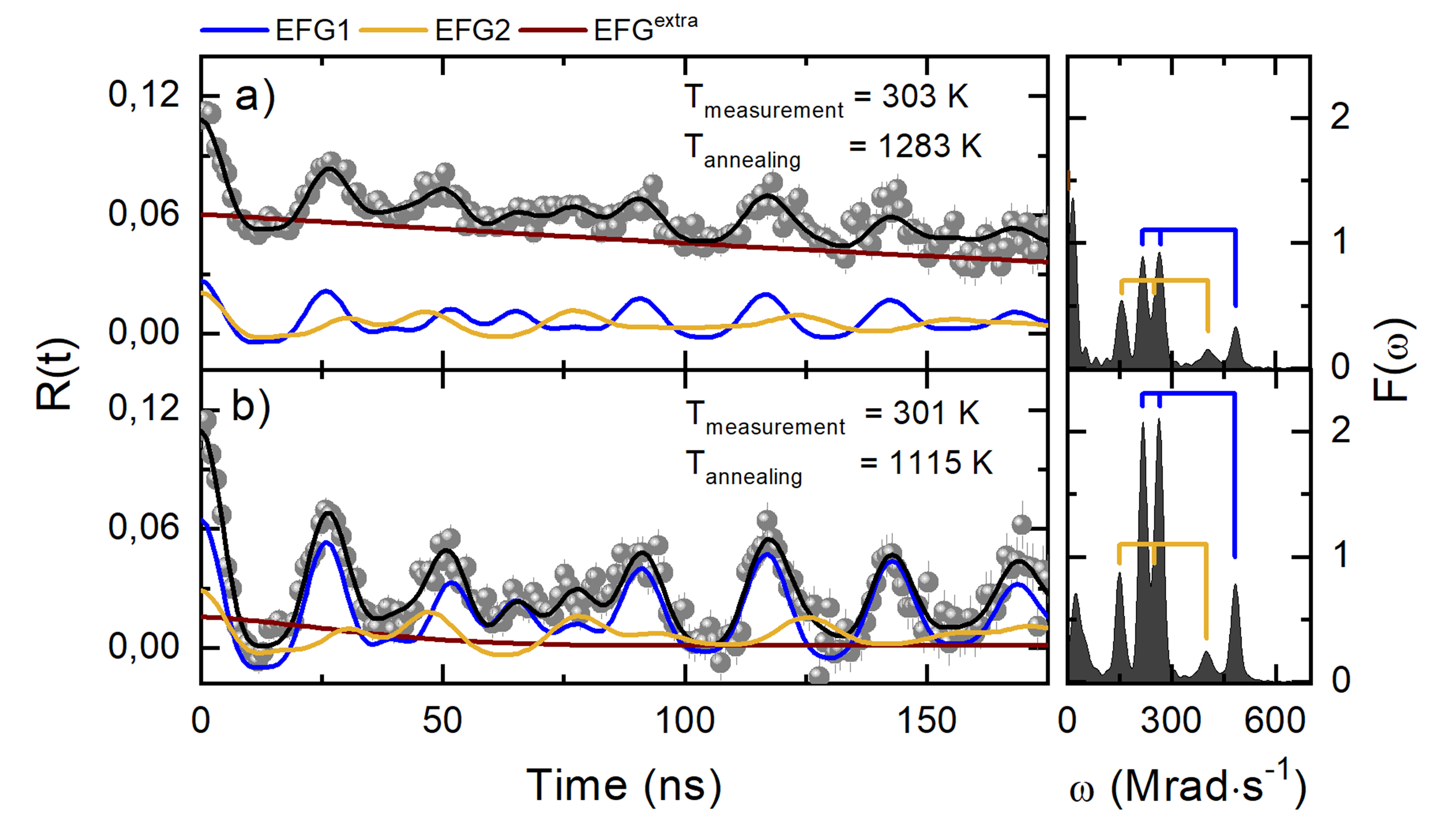}
	\caption{Comparing the effects of increasing the post-implantation annealing temperature (T$_{\rm{annealing}}$) from 1115~K to 1283~K, by measuring afterward the respective $R(t)$ function near room temperature conditions: a) ${\rm T_{m}}=301$~K and b) ${\rm T_{m}}=303$~K. The corresponding fits are shown as the black lines and the individual contributions of the EFG$_1$, EFG$_2$, and EFG$^{\rm{extra}}$ distributions to the fits are also shown by the blue, yellow, and brown lines. The respective Fourier transforms of the fits are also displayed.}
	\label{fig:CTORT_appendix}
\end{figure}

An important observed phenomenon in fitting the experimental results is the need to add an extra phase associated with an extra EFG distribution of low strength and relative high line broadening. In the perturbation functions represented in Fig.~\ref{fig:CTORT_appendix}, this  EFG distribution effect is highlighted by the damped and gradual time-changing individual contribution to $R(t)$ represented by the brown lines (labeled as EGF$^{\rm{extra}}$).
The probe weight of this distribution ($f_i$) depends on the post-implantation annealing (T$_{\rm{annealing}}$) or the PAC measurement temperature (T$_{\rm{m}}$). Particularly, this annealing temperature effect is highlighted in Fig.~\ref{fig:CTORT_appendix}, where the experimental $R(t)$ functions were measured for identical temperatures ($\approx$ 300~K), but the post-implantation annealing was performed at 1283~K and 1115~K, respectively. The range of these annealing temperatures crosses the 1150~K value, for which decomposition of the \ch{Ca3Ti2O7} structure was reported to be noticeable in high-temperature XRD measurements~\cite{Senn_PRL2015}. The respective probe occupation of this low-strength EFG distribution increased from 23\% up to 58\%. Considering the case of static line broadening, such EFG distribution represents the probe fraction subjected to a large number of defects, where the average effect leads to a damped $R(t)$ partial function. The nature of this field distribution may result from combined effects of the defects induced in the \ch{^{111m}Cd} beam implantation process and of a local \ch{Ca3Ti2O7} partial decomposition under high-temperature conditions. The thermal stability of the \ch{Ca3Ti2O7} and \ch{Ca3Mn2O7} systems, including the stability limit of the NLP structures to the degree of Ca by Cd substitution, has been studied previously. In this previous DFT-based work~\cite{Marcondes2021}, it was observed that the Ti based-system is more prone to decomposition than the Mn-based one, into the respective \ch{CaO} and parent perovskite phases of \ch{CaTiO3} and \ch{CaMnO3}. Particularly, the \ch{CaO} phase as described in the cubic $Fm\bar{3}m$ space group, is expected to have null $|V_{zz}|$ at the calcium nuclear sites and could be one of the sources of this low-strength EFG-distribution. However, the exact nature of this damped distribution is beyond the scope of this work.

\end{document}